\documentstyle[12pt]{article}
\begin{document}

\begin{titlepage}
\begin{flushright}
       {\bf UK/95-01}  \\
 Jan. 1995   \\
       hep-ph/9502334
\end{flushright}
\begin{center}

{\bf {\LARGE Flavor-Singlet $g_A$ from Lattice QCD}}

\vspace{1cm}

{\bf S.J. Dong, J.-F. Laga\"{e}, and K.F. Liu   } \\[0.5em]
 {\it  Dept. of Physics and Astronomy  \\
  Univ. of Kentucky, Lexington, KY 40506}

\end{center}

\vspace{0.4cm}

\begin{abstract}

We report on a lattice QCD
calculation of the flavor-singlet axial coupling constant
$g_A$ of the proton from the axial-vector current.
The simulation is carried out at $\beta = 6$ on a quenched $16^3
\times 24$ lattice. An extrapolation to the chiral limit
 shows that the connected insertion (valence and cloud parts)
is $0.62 \pm 0.09$, which is close to $g_A^8$. The disconnected
insertion (vacuum polarization from the sea quark) for the $u$ or $d$
quark is $- 0.12\pm 0.01 $ and the $s$ quark also
contributes $-0.12\pm 0.01$. The total $g_A^1$ is thus $0.25 \pm 0.12$
which is in agreement with experiments. In addition, we find
$g_A^3 = 1.20 \pm 0.11$, $g_A^8 = 0.61 \pm 0.13$
 and $F_A/D_A = 0.60 \pm 0.02$, also in agreement with experiments.
\bigskip


\end{abstract}

\vfill

\end{titlepage}

Recent experiments on polarized deep inelastic lepton-nucleon
scattering from SMC~\cite{smc94} and E143~\cite{e14395} have
confirmed the finding of the earlier EMC~\cite{emc88} results that
the flavor-singlet $g_A^1$ is small~\cite{bek88}.
In the deep inelastic limit, the integral of the polarized
structure function is related to the forward matrix elements of
axial currents from the operator product expansion~\cite{kod80}.
Combined with the neutron and hyperon decays, the
flavor-singlet axial coupling $g_A^1$ is extracted. Since the
axial current is the canonical spin operator, $g_A^1$ is thus
the quark spin content of the nucleon; i.e.
$g_A^1 = \Delta u + \Delta d + \Delta s$,
where the spin content $\Delta q (q = u,d,s)$ is
defined in the forward matrix element of the axial current,
 $\langle ps|\bar{q}i\gamma_{\mu}\gamma_5 q | ps \rangle
= 2 M_N s_{\mu} \Delta q$.

The fact that $g_A^1$, which represents the quark
spin contribution to the proton spin, is found to be much smaller
than the expected value of unity from the
non-relativistic quark model or 0.75 from the SU(6) relation (
3/5 of the isovector coupling $g_A^3 = 1.2574$)
has attracted a lot of attention. Despite numerous
attempts~\cite{bt93} to explain the smallness of $g_A^1$
by means of various hadronic models, effective theories,
and anomalous Ward identity~\cite{liu92},
this problem has persistently defied
satisfactory answer and has been dubbed the ``proton spin crisis''.

Attempts have also been made to calculate $g_A^1$ via
the anomalous Ward identity using lattice QCD, which is well poised
for such a challenge \cite{gm94,agh94}.
However, it has been pointed out \cite{gm94,liu92} that in the quenched
approximation, the induced
pseudoscalar coupling can not be ignored due
to the would-be Goldstone boson dominance in the forward matrix
element. As a result, the topological coupling to the nucleon
$\langle N|G\tilde{G}|N\rangle$ is not directly related to $g_A^1$
in the quenched approximation. Lattice calculations with dynamical
fermions are free of this problem. Modulo uncertainties from the
chiral and zero momentum extrapolation, a recent calculation of the
topological coupling of the nucleon using dynamical staggered fermions
\cite{agh94}
shows that the quark spin content is indeed small, comparable
to the experimental findings.
However, a calculation
of $g_A^1$ via the Ward identity does not address the issue why it
is much smaller than that of the quark model, a question which is
at the heart of the spin crisis. In an attempt
to answer this question, we carry out a lattice calculation of
the axial current directly.
It turns out that the polarization due to the degrees of freedom
outside of the naive quark model, i.e. cloud and sea quarks, are
responsible for the smallness of $g_A^1$.

Lattice calculations of three-point functions have been used to
study the EM~\cite{dwl90}, axial (isovector)
\cite{ldd94a}, and pseudoscalar($\pi NN$) \cite{ldd94b} form factors
of the nucleon. For the flavor-singlet $g_A^1$, we calculate
the following two- and three-point functions,

\begin{equation} \label{twopt}
G_{PP}^{\alpha\alpha}(t        ) = \sum_{\vec{x}}
          \langle 0|    \chi^\alpha(x) \bar{\chi}^\alpha(0) |0 \rangle
\end{equation}
\begin{equation} \label{threept}
\Gamma_3^{\beta\alpha} G_{PA_3P}^{\alpha\beta}(t_f, t)=
\sum_{\vec{x}_f,\vec{x}} \Gamma_3^{\beta\alpha} \langle 0|
\chi^\alpha(x_f) A_3(x)   \bar{\chi}^\beta(0)  |0 \rangle ,
\end{equation}
where $\chi^\alpha$ is the proton interpolating field, $\Gamma_3
= -i \gamma_3\gamma_5 (1 + \gamma_4)/2$, and $A_3$
is the point-split axial current from the Wilson action
\begin{equation} \label{psc}
A_\mu      \!=\! 2\kappa f(ma)
[\bar{\psi}(x)\frac{1}{2}i \gamma_\mu\gamma_5
U_\mu(x) \psi(x+\hat{\mu}) \!+\! \bar{\psi}(x+\hat{\mu})
\frac{1}{2}i \gamma_\mu\gamma_5 U_\mu^\dagger (x)\psi(x)].
\end{equation}

   In contrast to the isovector case, the evaluation of the
 three-point function for the flavor-singlet
current involves a {\it disconnected insertion}(DI) in addition
to the {\it connected insertion}(CI) \cite{liu92}.
The quark line skeleton diagrams
for the CI and DI are shown in Fig. 1(a) and 1(b) respectively. The
DI refers only to the quark lines. They are
nonetheless correlated via the background gauge fields.
We shall calculate the CI and DI separately. The factor $f(ma)$
in eq. (\ref{psc}) is the finite
$ma$ correction for the Wilson action. We take $f(ma) = e^{ma} =
 (4\kappa_c/\kappa -3)$ for the connected insertion. For the
disconnected insertion, the factor $f(ma)$ is different from
$e^{ma}$ and depends on the Lorentz structure of the current.
We shall use the value of $f(ma)$ computed by comparing the
   triangle diagram in the lattice Wilson action and the
continuum \cite{ll95}.

The connected insertion is calculated in
the same way as the isovector
axial coupling $g_A^3$ \cite{ldd94a} where eqs. (\ref{twopt}) and
(\ref{threept}) are fitted to two exponentials in the form
$fe^{-mt_f'}$ and $g_{A, con}^{1L}fe^{-mt_f}$ simultaneously, using the
data-covariance matrix to account for correlations. In evaluating
the three-point function,
the factor $8\kappa_c \langle \frac{1}{3}
TrU_{plaq}\rangle^{1/4}$ ($\kappa_c = 0.1568$)
is divided from the axial current in eq.
(\ref{psc}) to account for the mean-field improvement of
the lattice operator \cite{lm93,ldd94a,ldd94b}.
Numerical details are given in Ref. \cite{ldd94a}.
The unrenormalized lattice
$g_{A, con}^{1L}$ has been calculated for $\kappa = 0.154, 0.152$,
and 0.148,
corresponding to quark masses of about 120, 200, and 370 MeV
respectively (the scale $a^{-1} = 1.74(10)$GeV is set by the nucleon
mass), and is plotted in Fig. 2. The calculations were done on a
quenched $16^3 \times 24$ lattice at $\beta = 6.0$ with 24 gauge
configurations as in the isovector case \cite{ldd94a}. Extrapolation
to the chiral limit ($\kappa_c = 0.1568$) yields $g_{A, con}^{1L}
= 0.65 \pm 0.09$ as shown in Fig. 2. The $g_A^1$ in the continuum
is related to its lattice counterpart by the relation $g_A^1
= Z_A g_A^{1L}$, where $Z_A$ is the finite lattice renormalization
constant. The one-loop calculation gives $Z_A = 0.952$ for
$\beta = 6.0$~\cite{lm93}, from which we find
$g_{A, con}^1 = 0.62 \pm 0.09$.

First we note that $g_{A, con}^1 = \Delta u_{con}
+ \Delta d_{con}$
is the OZI preserving part of $g_A^1$. If the DI
part (sea-quark contribution) is roughly flavor-independent, i.e.
$\Delta u_{dis} = \Delta d_{dis}  \simeq \Delta s$ (strange quarks
appear only in DI), as will be shown below,
 then $g_{A, con}^1$ should be close to $g_A^8$, the octet axial
coupling, i.e.
\begin{equation}
g_A^8 = \Delta u_{con} +
\Delta d_{con} + (\Delta u_{dis} + \Delta d_{dis} - 2\Delta s)
\simeq g_{A, con}^1
\end{equation}
{}From the recent fit of the nucleon and hyperon $\beta$ decays,
$g_A^8 = 0.579 \pm 0.021$ \cite{cr93}.
We see that our calculated $g_{A, con}^1$
is quite in agreement with this. Second, we notice that the
calculated ratio $R_A = g_{A, con}^1/g_A^3$
is smaller than the expected value of 3/5 from the SU(6) relation
of the relativistic quark model. The behavior of
this ratio $R_A$ is plotted in Fig. 3 as a function of the quark mass.
For heavy quarks (at ma = 1, m = 1.74 GeV), $R_A$ is 3/5. This is to be
expected of the non-relativistic quarks where $g_A^3 = 5/3$ and $g_A^1
= 1$. For light quarks, $R_A$ becomes progressively less than 3/5. It
has been shown \cite{ld94a} that this deviation from the SU(6) relation
is due to the presence of cloud quarks and antiquarks (those in the
higher Fock space than the valence in the CI). When this degree of
freedom is eliminated by disallowing quarks from propagating backward
in time, it is found that $R_A$ becomes 3/5 (shown as dots in Fig. 3
where the errors are smaller than the dot size). The
SU(6) relation is therefore recovered in the
{\it valence approximation}, in which the Fock
space is limited to the valence quarks. We should mention that under
this valence approximation, the quark spin content is still not equal
to the
non-relativistic value of unity. This has been explained~\cite{jm90}
as a relativistic effect. Owing to the presence of the
lower component in the Dirac spinor, a confined quark with s-wave
upper component still has a non-vanishing orbital angular momentum.
Thus we conclude that the reason $g_{A, con}^1$ is only about 60\%
of the non-relativistic value of 1 is because of relativistic
effects and polarization due to cloud quarks and antiquarks.

   To calculate the DI in Fig. 1(b), we sum the ratio
between eq. (\ref{threept}) and eq. (\ref{twopt})
over t. It has been shown \cite{mmp87} that
as $t_f >> a$, this sum becomes
\begin{equation} \label{ratiodis}
\frac{1}{3} \sum_{i = 1}^3
\sum_t\frac{\Gamma_i^{\beta\alpha}G_{PA_iP}^{\alpha\beta}(t_f,t
       )} {G_{PP}^{\alpha\alpha}(t_f        )}\,\,
{}_{\stackrel{\longrightarrow}{t_f >>a}}\,\, {\rm const}
 + t_f g_{A, dis}^{1L}
\end{equation}

Thus, we calculate the sum as a function of $t_f$ and
take the slope to obtain the DI part of $g_A^1$. Since the DI
involves  quark loops, and thus entails the calculation of off-diagonal
traces of the inverse quark matrix, it poses a challenging numerical
problem, for the size of the quark matrix is as large as
$10^6 \times 10^6$ in
the present case. In \cite{dl94}, we developed an efficient algorithm
to estimate these traces stochastically with $Z_4$ noise,
which gives the estimate with the
minimum variance. This algorithm has been tested by computing
quantities for which the exact answers are known.
We have checked the cases of
DI with vector and pseudoscalar currents. In both of these cases,
the matrix elements are proportional to the 3-momentum transfer
$q$. Hence, the forward matrix elements should vanish.
Presented in Fig. 4 are results of the ratios in eq. (\ref{ratiodis})
for the corresponding currents. These are obtained with 300 $Z_4$
noise vectors in each of the 50 gauge configurations for $\kappa
= 0.148$. To avoid contaminations by the boundary effects
introduced by the fixed boundary condition we imposed in the time
direction, we summed $t$ from the nucleon source, which
is 4 time slices away from the boundary, to 4 time slices from
the other boundary. The slopes of the ratios (indicated as ME in
Fig. 4) are fitted from the point where the nucleon emerges as a
single exponential in the two-point function which is at $t = 8$ and
onward. We see that for both the vector and pseudoscalar cases,
the calculated forward m.e. are indeed consistent with zero. We also
calculated the scalar m.e. $\langle N|\bar{q}q|N\rangle_{dis}$.
We see that the slope is large and positive, and thus has the
right magnitude
and sign to remove the discrepancy between the $\pi N \sigma$ term
calculated from $\pi N$ scattering and those lattice calculations
with only the CI. Similarly, the corresponding results
for $\kappa = 0.154$ are shown in Fig. 4. This is the lightest quark
we considered and it yields the largest errors. However, the
forward m.e. for the vector and pseudoscalar currents are still
consistent with zero and the scalar m.e. is larger than that of
$\kappa = 0.148$. We conclude from this study that stochastic
estimation with $Z_4$ noise produces the correct results
within errors.

    Having tested our algorithm against known quantities, we
proceed to calculate the axial-vector current. The results for
$\kappa = 0.148,
0.152$ and 0.154 are presented in Fig. 5. They are obtained
in the following way. First, the slopes are fitted in the region
$t_f \ge 8$ where the nucleon is isolated from its excited states.
The fit employs a data-covariant matrix to take into account
the correlation among the time slices in
the 50 gauge configurations and is fitted over
different ranges of $t_f$ to find the one with the minimum $\chi^2$,
much in the same way the hadron masses are fitted. The $\chi^2$ per
degree of freedom is small in each of the 3 cases considered here. They
are given in Fig. 5. The resultant fits covering the ranges of
$t_f$ with the minimum $\chi^2$ are plotted in Fig. 5.
Finally, the errors on the fit, also shown in the figure, are
obtained by jackknifing the procedure. We see that the slopes
are negative in the case of the axial current, in marked contrast to
the scalar current in Fig. 4.

In order to compare our results with experiments, we perform the
extrapolation to the chiral limit. Plotted in Fig. 6 are the results
of $g_{A, dis}^1$ for a single flavor with the same
sea-quark mass (denoted as $\kappa_1$) as those of the
valence- (and cloud-) quarks in the nucleon ($\kappa_2$). These results
include the one-loop renormalization constant $Z_A$ as in the
CI. The DI has a two-loop log divergence in the continuum from the
triangle diagram insertion on a quark line \cite{kod80}.
We computed the finite lattice renormalization associated with this
2-loop contribution
and found it to be much smaller than the one-loop
result. Hence, we have neglected it here. The extrapolation is
done with the covariant matrix to consider the correlation
among the 3 $\kappa's$. The error on the chiral limit
result is again obtained by jackknifing the procedure of the
extrapolation. To calculate $\Delta s$ (the strangeness contribution
to $g_{A, dis}^1$, we fix $\kappa_1$ (sea-quark mass) at 0.154
and extrapolate $\kappa_2$ (valence-quark mass)
to the chiral limit. These results are plotted in Fig. 7.

    From Fig. 6 and Fig. 7, we find that $\Delta u_{dis} =
\Delta d_{dis} = - 0.12 \pm 0.01$ and $\Delta_s = - 0.12 \pm 0.01$.
Together, we obtain $g_{A, dis}^1 = - 0.36 \pm 0.03$. Combined with
$g_{A, con}^1$, we finally obtain $g_A^1 = 0.25 \pm 0.12$ which is in
good agreement with experiments \cite{smc94,e14395}. We tabulate
these and other results in Table 1 and compare with experiments.
We find that they all are in good agreement with experiments.
We should mention that our calculation on $g_A^1$ is in agreement
with a recent similar calculation with the volume source \cite{fko95}
which predicts $g_A^1 = 0.18(10)$. However, their $g_A^3$, $g_A^8$,
$F_A$, and $D_A$ are smaller than ours and the experiments by
$\sim 20\%$.
This presumably is attributable to the fact that their lattice
spacing at $\beta = 5.7$ is about $45\%$ larger than ours at
$\beta = 6.0$. We would like to point out an interesting observation.
Comparing Fig. 6 ($\kappa_1 = \kappa_2$) and Fig. 7 ($\kappa_1$
fixed at 0.154, the strange quark mass), we notice that although
the sea-quark mass (related to $\kappa_1$) changes by a
factor of 3 in Fig. 6, the results still coincide with those
in Fig. 7 for each of the valence-quark
case. It shows that the DI depends sensitively on the valence
quark mass but is independent of the sea-quark mass in the loop within
errors. This
is reminiscent of the finding that, after the finite $ma$ correction
for the Wilson action, the coupling between would-be U(1)
Goldstone bosons is independent of the mass in the two $\gamma_5$
loops~\cite{ll95} for light quarks. The two currents are related
to each other via the anomalous Ward identity and, in this context,
the mass independence is consistent with a DI of $\gamma_5$ current
being dominated by the zero modes. It would certainly be interesting
to try to verify this with an explicit calculation of the zero
modes.

 \begin{table}[ht]
\caption{Axial coupling constants and quark spin contents of proton in
comparison with experiments}
\begin{tabular}{llll}
 \multicolumn{1}{c}{} &\multicolumn{1}{c}{This Work} &
 \multicolumn{2}{c} {Experiments} \\
 \hline
 $g_A^1 = \Delta u + \Delta d + \Delta s$ & 0.25(12)& 0.22(10)
 \cite{smc94} & 0.27(10)\cite{e14395}  \\
 $g_A^3 = \Delta u - \Delta d$ & 1.20(10) \cite{ldd94a} & 1.2573(28)& \\
 $g_A^8 = \Delta u + \Delta d - 2\Delta s$ & 0.61(13) &
 0.579(25) \cite{cr93}  & \\
 $\Delta u $ & 0.79(11) & 0.80(6)\cite{smc94} & 0.82(6)\cite{e14395} \\
 $\Delta d $ & - 0.42(11) &  - 0.46(6)\cite{smc94} & - 0.44(6)
 \cite{e14395}  \\
 $\Delta s $ & - 0.12(1) & - 0.12(4)\cite{smc94} & - 0.10(4)
 \cite{e14395}  \\
 $F_A = (\Delta u - \Delta s)/2$ & 0.45(6) & 0.459(8) \cite{cr93} & \\
 $D_A = (\Delta u - 2 \Delta d + \Delta s)/2$ & 0.75(11) & 0.798(8)
 \cite{cr93} & \\
 $F_A/D_A$ & 0.60(2) & 0.575(16) \cite{cr93} &  \\
 \hline
 \end{tabular}
 \end{table}

In summary, we have computed both the connected and disconnected
insertions of the axial current in the proton in a quenched lattice
calculation. Albeit they all agree with experiments, the systematic
errors due to finite volume, discretization, renormalization,
and quenched approximation (which could
be as large as $7\%$ -- $20\%$\cite{ldd94b}) will have to be addressed
in the future.
Nevertheless, the physical picture of $g_A^1$ is getting clearer.
The smallness of the quark spin content compared to
the non-relativistic value of unity is, first of all, due to
the fact that the combined relativistic effect and polarization
of the cloud-quarks reduces the CI to $0.62 \pm 0.09$, a value very
close to $g_A^8$, i.e. $g_{A, con}^1 \simeq g_A^8$. This is because
the DI is
almost independent of the flavors $u,d$, and $s$.  Furthermore, the
sea-quark polarization is large and in the opposite direction of
the proton spin. It is the sum of all these effects that produces
a small $g_A^1$. We should stress that our calculation is
gauge-invariant in that no gauge fixing is applied. Since there is
no gauge-invariant dimension-three axial operator for the non-abelian
gauge field \cite{cm83,jm90}, our result is not mixed with the
gluon spin. Only the quark spin content contributes. In this paper,
 we have not attempted
to determine the complete composition of the proton spin. That
would entail calculations of the orbital angular momentum and the
gluon spin, which we leave for a future investigation.

This work is partially supported by DOE
Grant DE-FG05-84ER40154. The authors wish to thank
T. Draper, C. McNeile, A. Shapere, and C. Thron for helpful comments.

{\bf Figure Captions}

\noindent

\noindent
Fig. 1 (a) The connected insertion. (b) The disconnected insertion.

\noindent
Fig. 2 The lattice $g_{A, con}^{1L}$ for the connected
insertion as a function of the quark mass $ma$. The chiral limit
result is indicated by $\bullet$.

\noindent
Fig. 3 The ratio $R_A = g_A^1/g_A^3$ for the CI is plotted as a
function of $ma$. The results of the valence approximation are shown
as $\bullet$.

\noindent
Fig. 4 The ratios in eq.({\ref{ratiodis}) are plotted for the vector,
pseudoscalar, and scalar currents with quark masses $\kappa =
0.148$ and 0.154. ME gives the fitted slope.

\noindent
Fig. 5 The ratios of eq.({\ref{ratiodis}) for the axial current are
plotted for the 3 $\kappa$ cases. ME is the fitted slope.

\noindent
Fig. 6 The DI of the axial current as a function of $ma$. The quark
masses in the valence and the sea are kept the same. The chiral limit
result is indicated by $\bullet$.

\noindent
Fig. 7 The same as in Fig. 6 except the sea quark mass
($\kappa_1$) is fixed at 0.154.

\begin{thebibliography}{99}
\bibitem{smc94}
D. Adams et al. (SMC), Phys. Lett. {\bf B329}, 399 (1994).

\bibitem{e14395}
K. Abe et al. (E143),  Phys. Rev. Lett. {\bf 74}, 346 (1995).

\bibitem{emc88}
J. Ashman et al. (EMC), Phys. Lett. {\bf B206}, 364 (1988).

\bibitem{bek88}
S.J. Brodsky, J. Ellis, and M. Karliner, Phys. Lett. {\bf B206},
309 (1988).

\bibitem{kod80}
J. Kodaira, Nucl. Phys. {\bf B165}, 129 (1980).

\bibitem{bt93}
For reviews, see for example S.D. Bass and A.W. Thomas, J. Phys.
{\bf G19}, 925 (1993); J. Ellis and M. Karliner, Talk at PANIC '93,
hep-ph/9310272; T.P. Cheng and L.F. Li, DPF Conf. '90, 569 (1990).

\bibitem{liu92}
K.F. Liu, Phys. Lett. {\bf B281}, 141 (1992).

\bibitem{gm94}
R. Gupta and J.E. Mandula, Phys. Rev. {\bf D50}, 6931 (1994).

\bibitem{agh94}
R. Altmeyer, M. G\"{o}ckler, R. Horsley, E. Laermann, and
G. Schierholz, Phys. Rev. {\bf D49}, R3087 (1994).

\bibitem{dwl90}
T. Draper,
R.M. Woloshyn and K.F. Liu, Phys. Lett. {\bf 234B}, 121 (1990);
W. Wilcox, T. Draper, and K.F. Liu, Phys. Rev. {\bf D 46}, 1109 (1992).

\bibitem{ldd94a}
K.F. Liu, S.J. Dong, T. Draper, J.M. Wu, and W. Wilcox,
Phys. Rev. {\bf D 49}, 4755 (1994).

\bibitem{ldd94b}
K.F. Liu, S.J. Dong, T. Draper, and W. Wilcox, Phys. Rev. Lett.
(to appear); UK/94-01, hep-lat/9406007.

\bibitem{lm93}
G.P. Lepage and P.B. Mackenzie, Phys. Rev. {\bf D48}, 2250 (1993).

\bibitem{ll95}
J.F. Laga\"{e} and K.F. Liu, UK/94-04, hep-lat/9501007.

\bibitem{cr93}
F.E. Close and R.G. Roberts, Phys. Lett. {\bf B316}, 165 (1993).

\bibitem{ld94a}
K.F. Liu and S.J. Dong, Phys. Rev. Lett. {\bf 72}, 1790 (1994).

\bibitem{jm90}
R.L. Jaffe and A. Manohar, Nucl. Phys. {\bf B337}, 509 (1990).

\bibitem{mmp87}
L. Maiani et al., Nucl. Phys. {\bf B293}, 420 (1987).

\bibitem{dl94}
S.J. Dong and K.F. Liu, Phys. Lett. {\bf B328}, 130 (1994).

\bibitem{fko95}
M. Fukugita, Y. Kuramashi, M. Okawa, and A. Ukawa,
KEK preprint 94-173, hep-lat/9501010.

\bibitem{cm83}
C. Cronstr\"{o}m and J. Michelsson, J. Math. Phys. {\bf 24}, 2528
(1983).

\end{thebibliography}
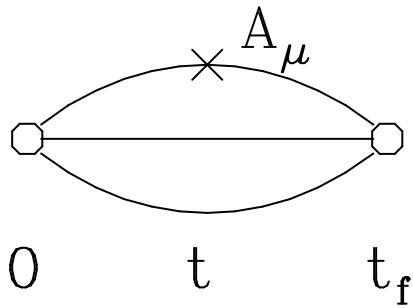
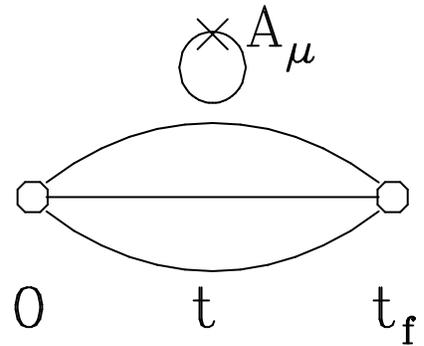
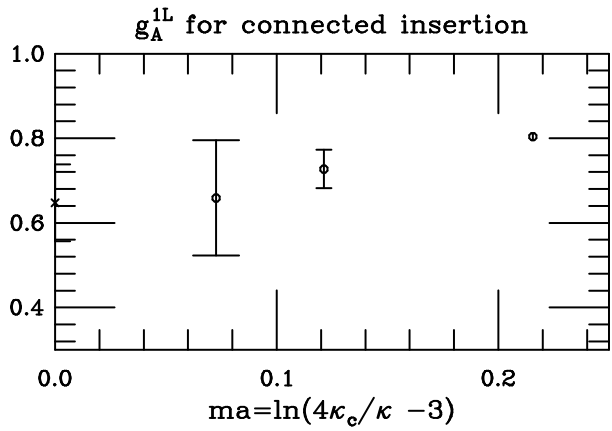
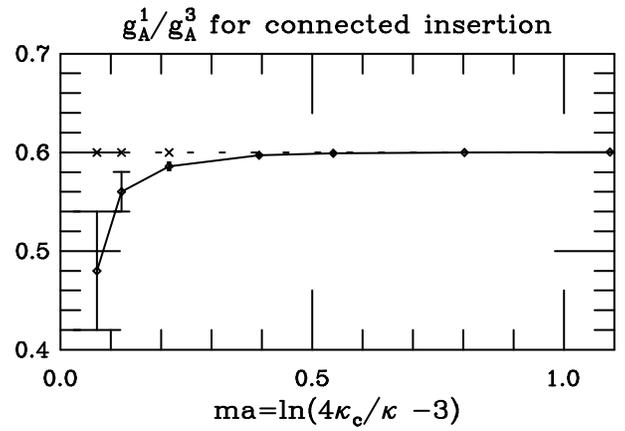
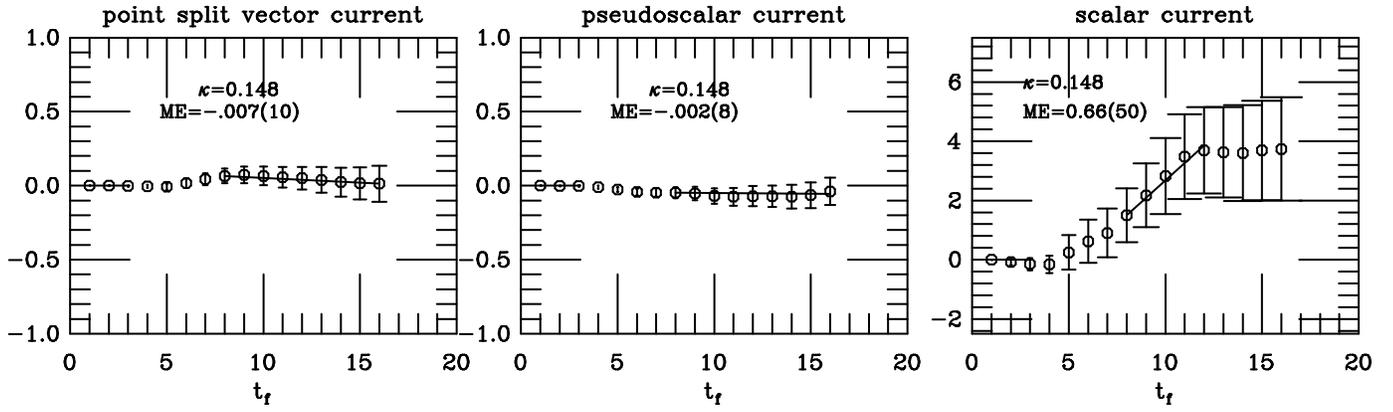
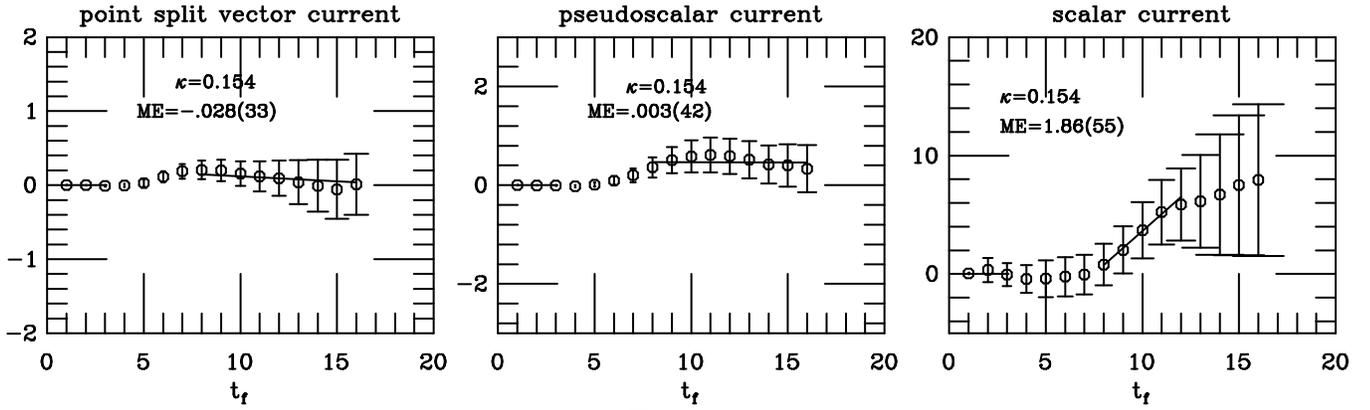
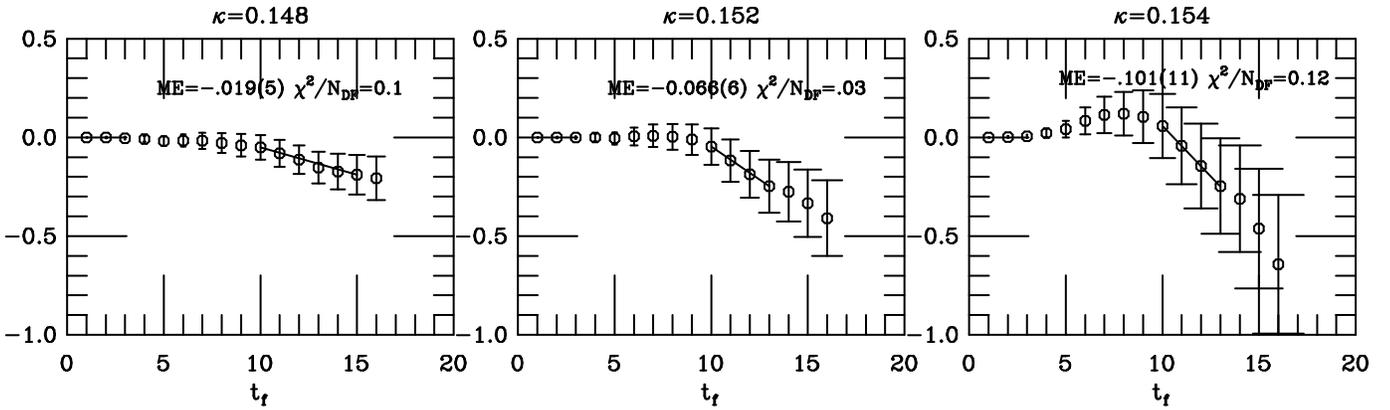
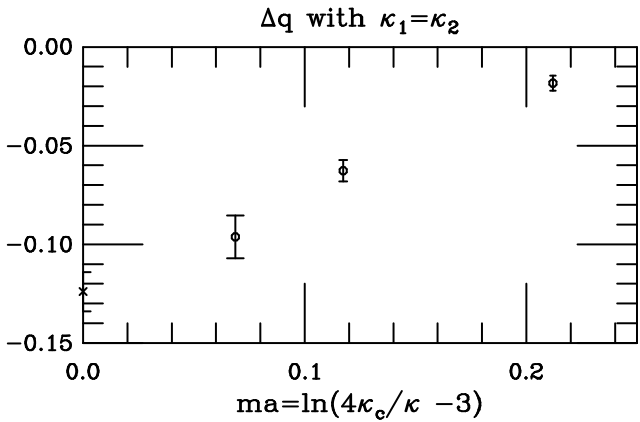
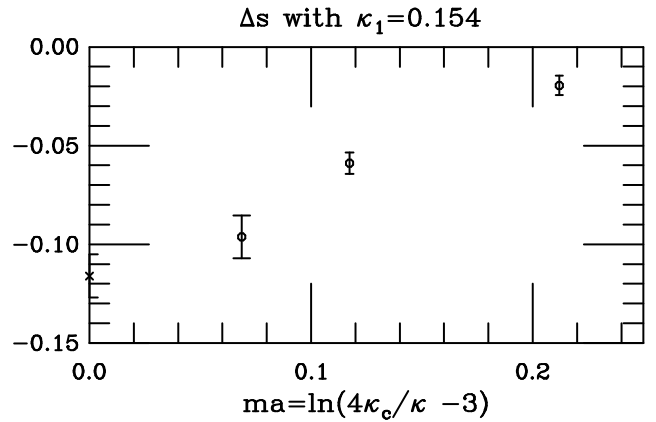
\end{document}